\documentclass{mem}
\usepackage{natbib}\usepackage{txfonts}\usepackage{balance}
\usepackage{graphicx}
\usepackage[a4paper,breaklinks,dvipdfm]{}
\idline{75}{282}
\begin{document}

\def\teff{$T\rm_{eff }$}
\def\kms{$\mathrm {km s}^{-1}$}

\newcommand{\DontShow}[1] {}
\newcommand{\ToDo}[1] {\textbf{ToDo: #1}}
\def\tens#1{\bar{\bar{\ensuremath{\mathsf{#1}}}}}
\newcommand{\papa}[2] { \frac{\partial #1}{\partial #2} }
\newcommand{\papaco}[3] {{ \left. \frac{\partial #1}{\partial #2}\right|_{#3}}}
\newcommand{\tildepapaco}[3] {{ \left. \widetilde{\frac{\partial #1}{\partial #2}}\right|_{#3}}}
\newcommand{\COBOLD}{CO5BOLD}
\newcommand{\I}[1] {i{\rm #1}}
\newcommand{\m}[1] {m{\rm #1}}
\newcommand{\n}[1] {n{\rm #1}}
\newcommand{\mghost}[1] {m{\rm #1}_{\rm ghost}}
\newcommand{\nghost}[1] {n{\rm #1}_{\rm ghost}}
\newcommand{\dx}[1] {\Delta x{\rm_#1}}
\newcommand{\xc}[1] {x_{\mathrm{c} #1}}
\newcommand{\xb}[1] {x_{\mathrm{b} #1}}
\newcommand{\g}[1]{g{\rm_#1}}
\newcommand{\phic}[1]{\Phi_{\mathrm{c} ,\,  #1}}
\newcommand{\phib}[1]{\Phi_{\mathrm{b} ,\,  #1}}
\newcommand{\V}[1]{v_{#1}}
\newcommand{\B}[1]{B_{#1}}
\newcommand{\tildeV}[1]{\tilde{v}{\rm #1}}
\newcommand{\Press} {P}
\newcommand{\ei} {e_\mathrm{i}}
\newcommand{\eg} {e{\rm g}}
\newcommand{\eikb} {e{\rm ikb}}
\newcommand{\eikbg} {e{\rm ikbg}}
\newcommand{\etot} {e_\mathrm{t}}
 \newcommand{\tildeei} {\tilde{e}{\rm i}}
\newcommand{\tildeeik} {\tilde{e}{\rm ik}}
\newcommand{\tildeek} {\tilde{e}{\rm k}}
\newcommand{\tildeeip} {\tilde{e}{\rm ip}}
\newcommand{\eikp} {e{\rm ikp}}
\newcommand{\tildeeikp} {\tilde{e}{\rm ikp}}
\newcommand{\dPdei} {\papaco{P}{\ei}{\rho}}
\newcommand{\dPdrho} {\papaco{P}{\rho}{\ei}}
\newcommand{\drhoeidP} {\papaco{\rhoei}{P}{s}}
\newcommand{\drhoeidrho} {\papaco{\rhoei}{\rho}{P}}
\newcommand{\tildedrhoeidP} {\tildepapaco{\rhoei}{P}{s}}
\newcommand{\rhov}[1]{\rho v{\rm_#1}}
\newcommand{\rhoei} {\rho\ei}
\newcommand{\rhoeik} {\rho e{\rm ik}}
\newcommand{\rhoeip} {\rho e{\rm ip}}
\newcommand{\rhoek} {\rho e{\rm k}}
\newcommand{\rhoeg} {\rho e{\rm g}}
\newcommand{\rhoeikg} {\rho e{\rm ikg}}
\newcommand{\CCour} {C_{\rm Courant}}
\newcommand{\CCourmax} {C_{\rm Courant,max}}
\newcommand{\iup} {i_{\rm up}}
\newcommand{\cs} {c_\mathrm{s}}
\newcommand{\vc} {v_\mathrm{c}}
\newcommand{\wL} {w{\rm L}}
\newcommand{\wR} {w{\rm R}}
\newcommand{\Pred}{{P_{\rm st}}}
\newcommand{\Deltat}{\mbox{\small $\Delta$} t}
\newcommand{\dvdx}[2] { \frac{\Delta \V{#1}}{\dx{#2}} }
\newcommand{\Source}{S}
\newcommand{\dSource}{\Delta \Source}

\newcommand{\Int}{I}
\newcommand{\dInt}{\Delta \Int}
\newcommand{\ImS}{\hat{I}}
\newcommand{\dImS}{\Delta \hat{I}}
\newcommand{\Ired}{\tilde{I}}
\newcommand{\dIred}{\Delta \tilde{I}}

\newcommand{\Dx}{\Delta x}

\newcommand{\dtau}{\Delta \tau}
\newcommand{\dedtau}[1] { \frac{{\rm d} #1}{{\rm d} \tau} }
\newcommand{\dedtaunu}[1] { \frac{{\rm d} #1}{{\rm d} \tau_{\nu}} }
\newcommand{\dededtau}[1] { \frac{{\rm d}^2 #1}{{\rm d} \tau^2} }
\newcommand{\dededtaunu}[1] { \frac{{\rm d}^2 #1}{{\rm d} \tau_{\nu}^2} }

\newcommand{\dedx}[1] { \frac{{\rm d} #1}{{\rm d} x} }

\newcommand{\elt} {\widehat{e}}

\newcommand{\wII}{w(2)}
\newcommand{\wIII}{w(3)}

\newcommand{\Snu}{\ensuremath{S_\nu}}
\newcommand{\Jnu}{\ensuremath{J_\nu}}
\newcommand{\taunu}{\ensuremath{\tau_\nu}}

\bibliographystyle{aa}

\title{Progress in Modeling Very Low Mass Stars, Brown Dwarfs, and Planetary Mass Objects}
 
 \subtitle{}

\author{
F. Allard, D. Homeier, B. Freytag, W. Schaffenberger, \& A. S. Rajpurohit
          }

  \offprints{F. Allard}

\institute{
CRAL, UMR 5574, CNRS,
Universit\'e de Lyon, 
\'Ecole Normale Sup\'erieure de Lyon, \\
46 All\'ee d'Italie, F-69364
Lyon Cedex 07, France,
\email{fallard@ens-lyon.fr}
}

\authorrunning{Allard et al. }

\titlerunning{Modeling the stellar-substellar transition}

\abstract{
We review recent advancements in modeling the stellar to substellar 
transition. The revised molecular opacities, solar oxygen abundances and cloud models 
allow to reproduce the photometric and spectroscopic properties of this transition to a 
degree never achieved before, but problems remain in the important M-L transition 
characteristic of the effective temperature range of characterizable exoplanets.  We 
discuss of the validity of these classical models. We also present new preliminary 
global Radiation HydroDynamical M dwarfs simulations.
}
\maketitle{}

\section{Introduction}
The spectral transition from very low mass stars (VLMs) to the latest type brown dwarfs is remarkable by the magnitude of the transformation 
of the spectral features over a small change in effective temperature.   It is characterized by i) the condensation onto seeds of strong opacity-bearing molecules such as CaH, TiO and VO which govern the entire visual to near-infrared part ($0.4 - 1.2~\mu$m) of the spectral energy distribution (hereafter SED); ii) a "veiling" by Rayleigh and Mie scattering of submicron to micron-sized aerosols; iii) a weakening of the infrared water vapor bands due to oxygen condensation and to the greenhouse (or blanketing effect) caused by silicate dust in the line forming regions; iv) methane and ammonia band formation in T and Y~dwarfs; and finally v) water vapor condensation in Y~dwarfs (\teff $~\le~ 500$~K).   Condensation begins to occur in M~dwarfs with 
\teff ~$< 3000$~K. In T~dwarfs the visual to red part of the SED is dominated by the wings of the Na\,I\,D and 0.77~$\mu$m K\,I alkali doublets which form out to as much as 2000\,\AA\  from the line center \citep{AllardN2007,Allard07}.  The SED of those dwarfs is therefore dominated by molecular opacities and resonance atomic transitions under pressure  ($\approx 3$~bars)  broadening conditions, leaving no window onto the continuum \citep[][1997]{AllardPhDT90}\nocite{ARAA97}.  
 
With the absence of magnetic breaking due to a neutral atmosphere \citep{Mohanty2002}, brown dwarfs should present differential rotation and clouds should be distributed in bands around their surface much as is shown for Jupiter. \cite{Marley2002} have suggested that brown dwarfs in the L-T transition are affected by cloud cover disruption.  And indeed, several objects show even large-scale photometric variability (Artigau et al.\ 2009, Radigan et al.\ 2012)\nocite{Artigau2009,Radigan2012}  --- on the order of 5\% to even 10-30\% in the best studied case. \cite{Buenzli2012} find periodic variability both in near-IR and mid-IR for a T6.5 brown dwarf in simultaneous observations conducted with HST and Spitzer. The phase of the variability varies considerably between wavelengths, suggesting a complex atmospheric structure. Recent large-scale surveys of brown dwarf variability with Spitzer (PI Metchev) have revealed mid-IR variability on order of a few percent in $>$~50\% of L and T type brown dwarfs. On the basis of these results, variability may be expected for young extrasolar planets, which share similar Teff and spectral types. And the ubiquity of cloud structures in L3-T8 dwarfs strongly suggests that these may persist into the cooler ($>$~T8) objects. 

The models developed for VLMs and brown dwarfs are a unique tool for the characterization of imaged exoplanets, if they can explain the stellar-substellar transition.  And global circulation models subjected to cloud formation in presence of rotation are necessary to explain the observed weathering phenomena. Recently, Allard et al. (2012a,b)\nocite{Allard2012a,Allard2012b} and \cite{LHS1070_2012} have published the preliminary results of a new model atmosphere grid computed with the \texttt{PHOENIX} atmosphere code accounting for cloud formation and mixing from Radiation HydroDynamical (RHD) simulations \citep{Freytag2010}. In this paper, we present the latest evolution in modeling their SED and their observed photometric variability and present new prospectives for this field of research.

\begin{figure*}[!ht]
  \includegraphics[width=14.5cm]{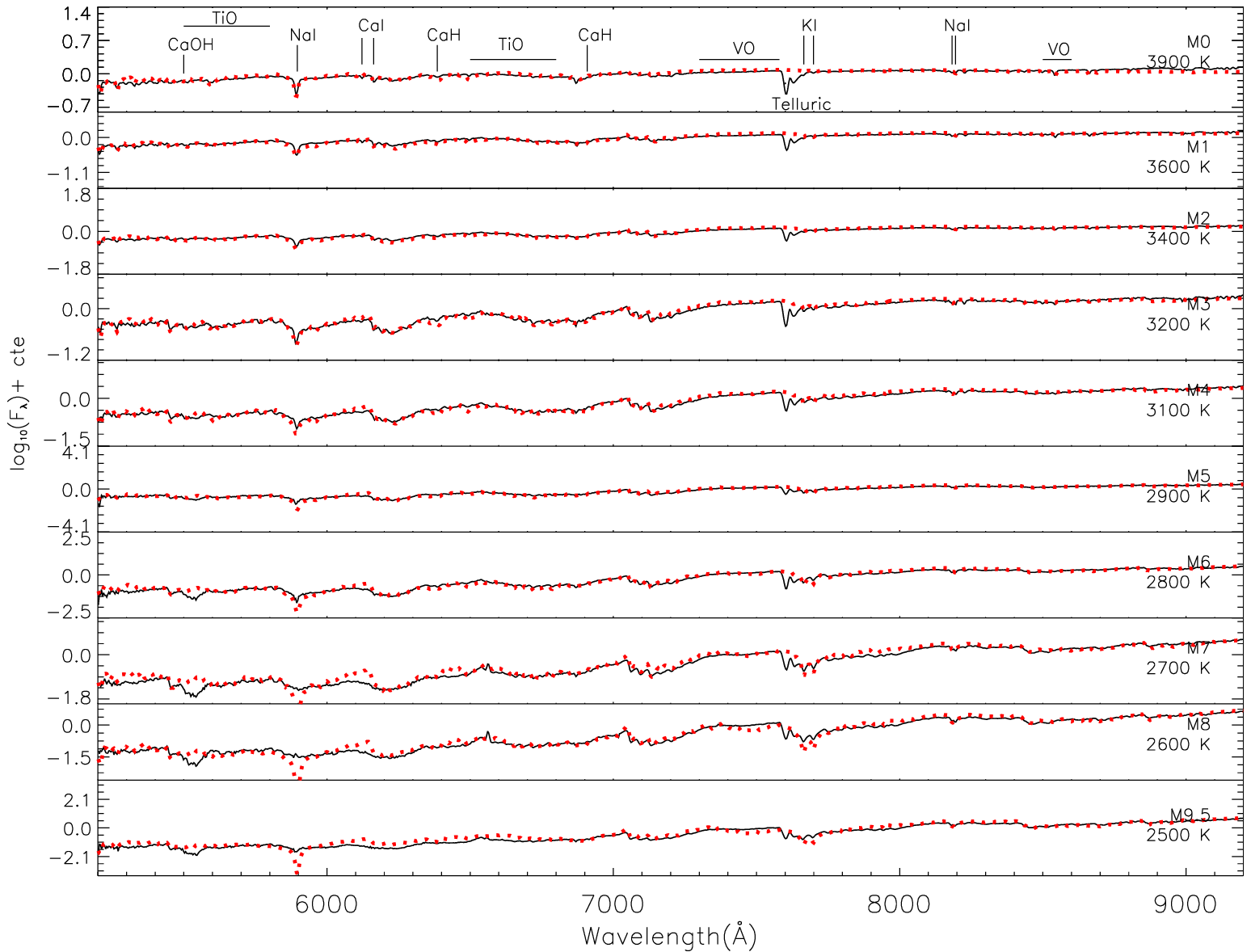}
  \caption{\footnotesize
    Fig. 1 of the article by Rajpurohit et al. (in prep.). The optical to red SED of M dwarfs from M0 to M9 observed with the NTT at a spectral  
    resolution of 10.4 \AA\ are compared to the the best fitting (chi-square minimization) BT-Settl synthetic spectra (dotted lines), assuming a solar 
    composition according to Caffau et al. (2011). The models displayed have a surface gravity of long=5.0 to 5.5 from top to bottom. The best fit is
    determined by a chi-square minimization technic. The slope of the SED is particularly well reproduced all through the M dwarfs spectral sequence.  
    However, some indications of missing opacities (mainly hydrides) persist to the blue of the late-type M dwarf cases such as missing opacities in the 
    B' $^{ 2}${$\Sigma$}$^{+}$$<$-- X $^{ 2}${$\Sigma$}$^{+}$ system of MgH by \cite{Story2003} and the opacities are totally missing 
    for the CaOH band near 5500\,\AA.  Note that chromospheric emission fils the Na\,I\,D transitions in the latest-type M dwarfs displayed here, 
    and that the M9.5 dwarf has a flatten optical spectrum due to dust scattering. 
    Telluric features near 7600\,\AA\ have been ignored from the chi-square minimization.}
  \label{f:Rajpurohit_fig1}
\end{figure*}

\section{M dwarfs}

\begin{figure*}[!ht]
  \includegraphics[width=13.0cm]{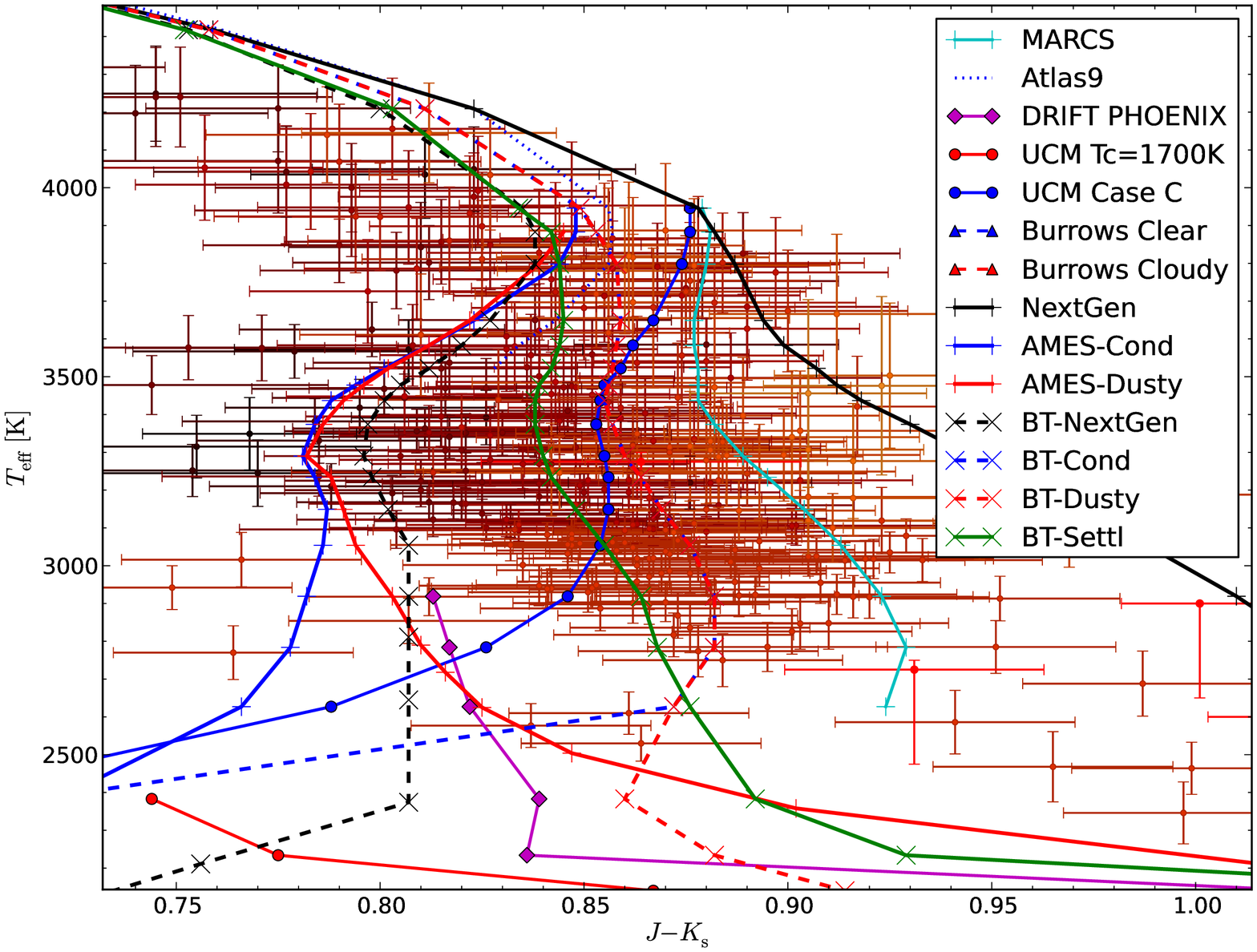}
  \caption{\footnotesize
   Estimated \teff\ and metallicity (decreasing from lighter to darker tones) for 
     M dwarfs by \cite{MdwarfsTeff2008} on the left, and brown dwarfs by Golimowski et al.\ (2004) 
     and Vrba et al.\ (2004) on the right are compared to the NextGen 
     isochrones for 5 Gyrs  \cite{BCAH98} using model atmospheres by various authors:  
     MARCS by \cite{MARCS2008}, ATLAS9 by \cite{ATLAS92004}, DRIFT-PHOENIX by 
     \cite{Helling08a}, UCM by \cite{tsuji02}, Clear/Cloudy by \cite{Burrows06}, NextGen 
     by \cite{NGa}, AMES-Cond/Dusty by \cite{Allard01},  
     the BT-Cond/Dusty/NextGen models by Allard et al.\ (2012)
     and the current BT-Settl models. 
     Some curves labeled in the legend can only be seen in the more extended Fig.~\ref{f:allard_f4_Teff-J-Ks}. }
  \label{f:allard_f2_Teff-J-Ks}
\end{figure*}

Because oxygen compounds dominate the opacities in the SED of VLMs, their synthetic spectra and colors respond sensibly to the abundance of oxygen 
assumed. Allard et al. (2012a,b)\nocite{Allard2012a,Allard2012b}  compare models using different solar abundance values \citep[][2007, Asplund et al. 2009]{GNS93}\nocite{Grevesse2007,Asplund09} and found an improved agreement with constraints \citep{MdwarfsTeff2008} for the solar abundances obtained using RHD simulations by Asplund et al. In this paper, we present models based on the \cite{Caffau2011} solar abundances. Taking the \cite{GNS93} solar abundance values as reference, the Grevesse et al. (2007) results show a reduced oxygen abundance by {-39\%}, while Asplund et al.  obtain a reduction by -34\%. These values poses problem for the interpretation of solar astero-se\"\i smological results \citep[][Antia \& Basu 2011]{Basu2008}\nocite{Antia2011}.  More recently, \cite{Caffau2011} used the CO5BOLD code to obtain a more conservative reduction of oxygen by -22\%. The latter value still allows an acceptable representation of the VLMs while preserving the astero-se\"\i smological solar results.   Higher spectral resolution and up-to-date opacities also contributed to improve VLMs models compared to previous versions \citep[][Allard et al. 2001]{NGa}\nocite{Allard01}. These models rely on lists of molecular transition determined ab initio. See the review by Homeier et al. elsewhere in this journal for the opacities used in the BT-Settl models presented in this paper.

The comparison of the BT-Settl \texttt{PHOENIX} models based on the \cite{Caffau2011} solar abundances to the low resolution spectra and to the \cite{Casagrande2008} temperature scale is shown in Figs.~\ref{f:Rajpurohit_fig1}  and \ref{f:allard_f2_Teff-J-Ks}.  Fig.~\ref{f:Rajpurohit_fig1} shows an unprecedented agreement with spectral type through the M dwarf spectral sequence of the models. One can see in Fig.~\ref{f:allard_f2_Teff-J-Ks} that the new BT-Settl models lie slightly to the blue of the BT-Dusty models by \cite{Allard2012a} based on the \cite{Asplund09} solar abundances, but largely to the red of the AMES-Cond/Dusty models by \cite{Allard01} and the BT-NextGen models \cite{Allard2012a} based on the \cite{GNS93} solar abundances. The even lower oxygen abundance values of \cite{Grevesse2007} cause  the MARCS models by \cite{MARCS2008} to lie to the right of the diagram. The NextGen models by \cite{NGa} also lie to the right of the diagram due, among others, to the missing and incomplete molecular opacities.

The BT-Settl models are computed solving the radiative transfer in spherical symmetry and the convective transfer using the Mixing Length Theory \citep[][see also Ludwig et al. 2002 for the exact formalism used in \texttt{PHOENIX}]{Boehm-Vitense58} using a mixing length as derived by the RHD convection simulations \citep[][Freytag et al. 2012]{Ludwig2002,Ludwig2006}\nocite{Freytag2012} and a radius as determined by interior models \citep[][2003, Chabrier et al. 2000]{BCAH98}\nocite{CBAH00,Baraffe03} as a function of the atmospheric parameters (\teff, surface gravity, and composition). 
\texttt{PHOENIX} use the classical approach consisting in neglecting the magnetic field, convective and/or rotational motions and other multi-dimensional aspects of the problem, and assuming that the averaged properties of stars can be approximated by modeling their properties radially (uni-dimensionally) and statically. Neglecting motions in modeling the photospheres of VLMs, brown dwarfs, and planets is acceptable since the convective velocity fluctuation effects on line broadening are hidden by the strong van der Waals broadening and the important molecular line overlapping prevailing in these atmospheres.   But this is not the case of the impact of the velocity fields on the cloud formation and wind processes (see section \ref{s:RHD_clouds} below).

\section{Global RHD M dwarf simulation}
\begin{figure}[!ht]
  \includegraphics[width=6.5cm]{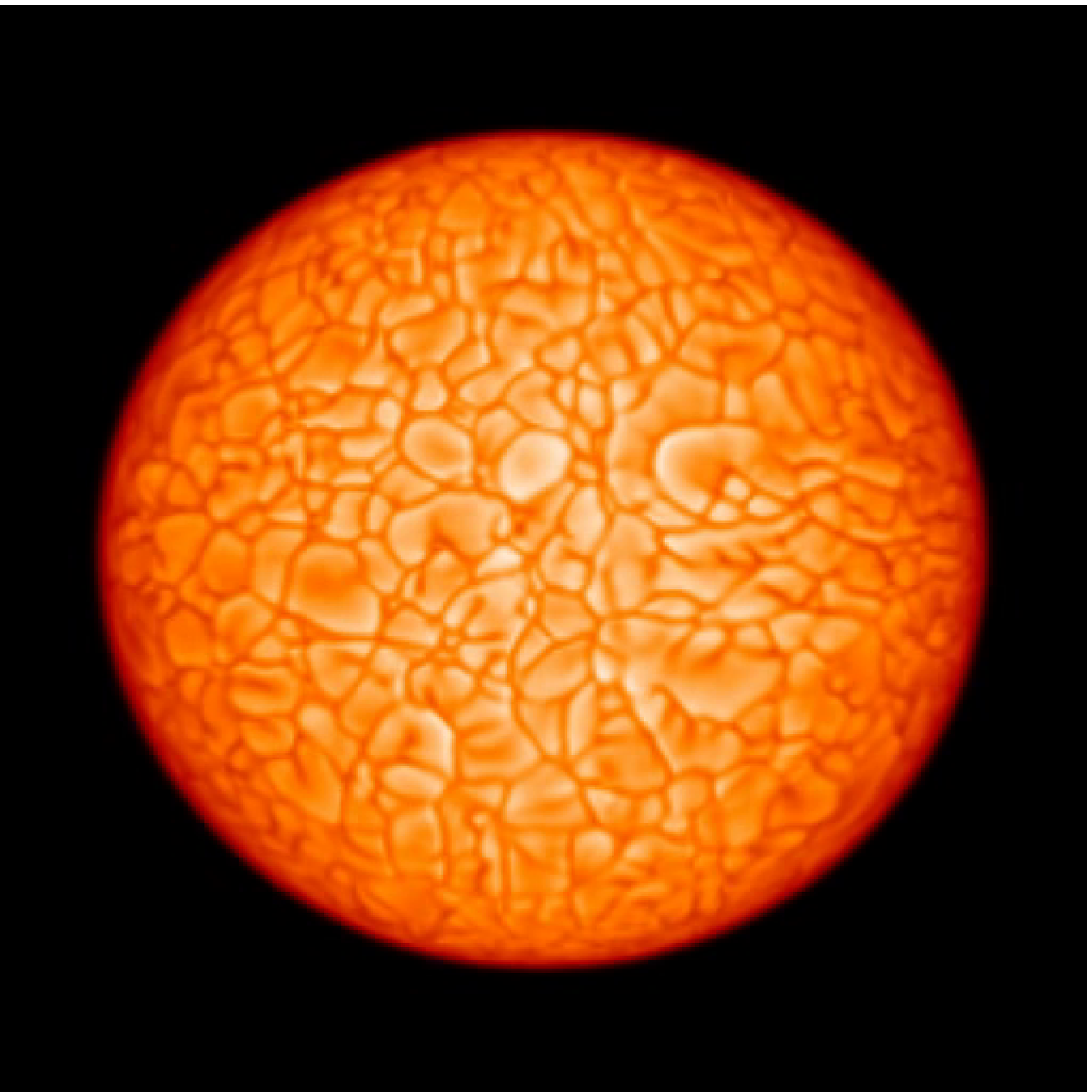}
  \caption{\footnotesize
   Global RHD simulation of an M dwarfs with solar composition. Obtained parameters are \teff $= 3143.44$\,K, log$g = 4.38$, and a radius of 0.003 R$_{\odot}$. The surface pressure is 68227.6 g/cm\,s$^2$, the surface density is 4.52146e-07 g/cm$^3$, the central temperature is 28889.7\,K, the central pressure is 1.67214e+09 g/cm\,s$^2$, the central density is 6.92703e-4 g/cm. The model covers 10 pressure scale heights from the center to the surface. The simulations use grey \texttt{PHOENIX} opacities in the outer layers merged with OPAL data for the interior layers. The rotation period of the simulation is 1~hour and the simulation is shown equator-on in the reference frame of rotation. One can notice a slight oblateness of the model at this exaggerated velocity for an M dwarf.  }
  \label{f:allard_f_fig3}
\end{figure}

VLMs and brown dwarfs are fully convective, and their convection zone extends into their atmosphere up to optical depths of even $10^{-3}$ \citep[][1997]{AllardPhDT90}\nocite{ARAA97}. Convection is efficient in M dwarfs and their atmosphere, except in the case of 1 Myr-old dwarfs which dissociate H$_2$, little sensitive to the choice of mixing length \citep{ARAA97}. 
A mixing length value between $\alpha =$ l/H$_p = 1.8 - 2.2$ has been determined by \cite{Ludwig2002,Ludwig2006} depending on surface gravity. These simulations have been recently extended to the late-type brown dwarf regime using CO5BOLD by \cite{Freytag2010}.  In brown dwarfs the internal convection zone retreats progressively to deeper layers with decreasing \teff.    

The magnetic breaking known to operate in low mass stars is not or less efficiently operating in fully convective M dwarfs ($<$ M3) and 
brown dwarfs. Brown dwarfs can therefore rotate with equatorial speeds as large as 60 km/sec or periods of typically 1.5 hour. 
Some have even been observed with periods nearly as low as their breakup velocity (1~hour). In comparison, planets of 
our solar system have a larger rotation period, such as 10 hours for Jupiter, and 24 hours for the Earth.  This rapid rotation 
(as well as their magnetic field) causes a suppression of the interior convection efficiency leading to a slowed down contraction 
during their evolution, and to larger radii then predicted by classical evolution models \citep{Chabrier2007}. 

\begin{figure*}[!ht]
  \includegraphics[width=13.0cm]{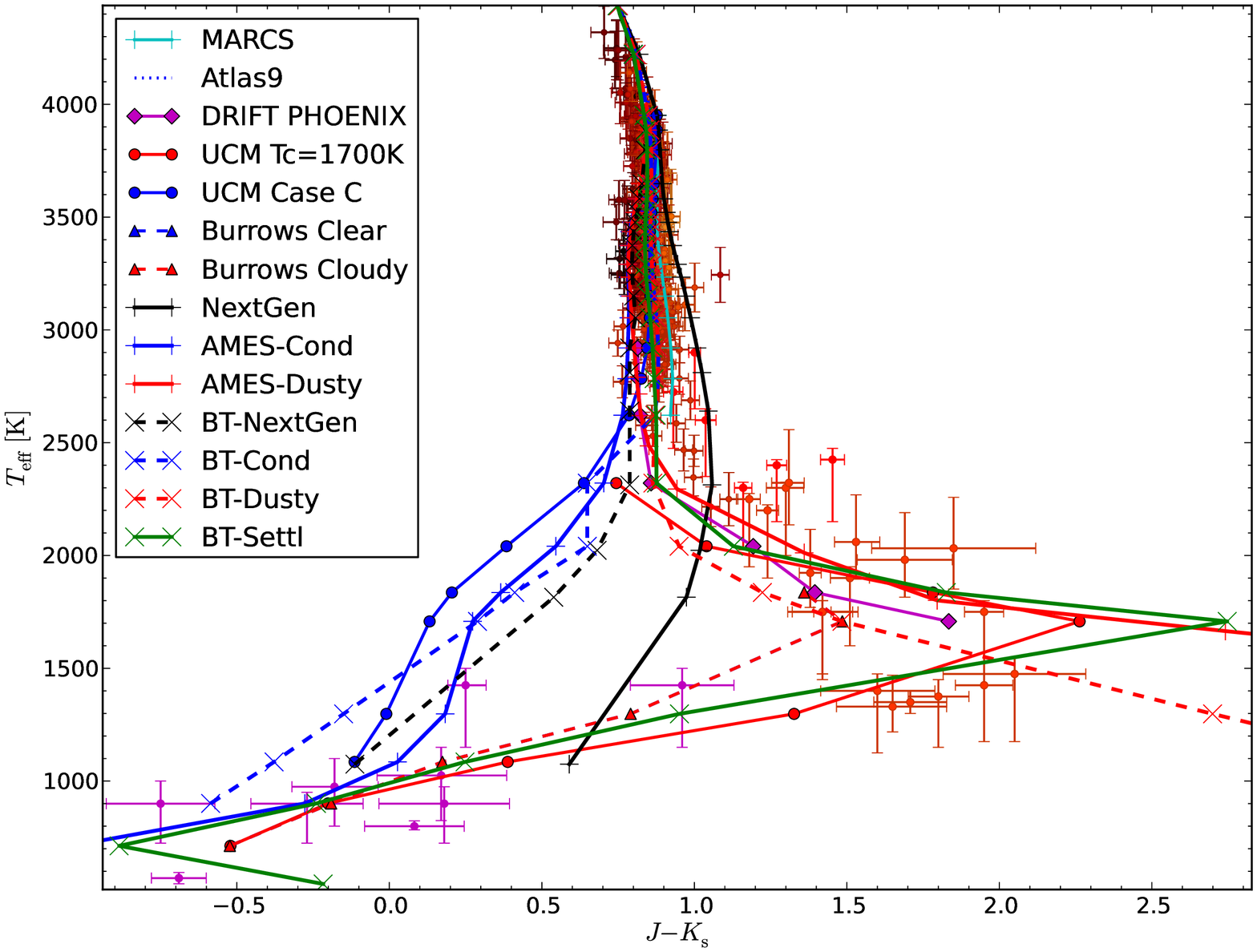}
  \caption{\footnotesize
   Same as Fig.~\ref{f:allard_f2_Teff-J-Ks} but extending into the brown dwarf regime for an age of 3 Gyrs.
    The region below 2900\,K is dominated by dust formation.  The dust free models occupy the 
     blue part of the diagram and only at best explain T dwarf colors, while the Dusty and DRIFT 
     models explain at best L dwarfs,  becoming only redder with decreasing {\teff}.  The BT-Settl, 
     Cloudy and UCM  $T_{\rm crit}=1700$\,K models describe a complete transition to the red in the 
     L dwarf regime before turning to the blue into the T dwarf regime, only the BT-Settl extends into the 
     Y dwarfs regime ({\teff} $\le  500$\,K).   }
  \label{f:allard_f4_Teff-J-Ks}
\end{figure*}

The CO5BOLD code solves the coupled equations of compressible hydrodynamics
and non-local radiative energy transport on a cartesian grid with a time-explicit scheme.
It can be used in a local setup (with constant downward gravity)
to model small patches of the stellar surface
or in a global ``star-in-a-box'' setup (with central gravitational potential)
to model entire stars.
The ``star-in-a-box'' setup has been used by \cite{Sun_Rotation07} to
compute global RHD simulations of the scaled-down Sun in
the presence of rotation, described by Coriolis and centrifugal forces.
In the current code version,
angular-momentum conservation and the isotropy of the flows
at small Mach numbers have been further improved.
This setup has been used, see Fig.~\ref{f:allard_f_fig3}, to compute a sequence of
radially scaled-down toy models of M dwarfs with various rotation periods
to determine the change in mixing length that corresponds to the
suppression of convection by rotation
(Schaffenberger \& Freytag, in prep.).
It is found that at realistic Coriolis numbers, rotation is indeed
able to reduce the convective flux (predominantly towards the equator)
to contribute significantly to the increase in radius
as suggested from analysing classical models.

\section{Cloud formation in \texttt{PHOENIX}}\label{s:BDs}

To describe the cloud formation in \texttt{PHOENIX} \cite{Allard2003} took the approach of using a cloud model drawn from an extensive study of cloud formation in planetary atmospheres \citep{Rossow1978} which compares layer-by-layer the timescales of the main processes (mixing, sedimentation, condensation, coalescence and coagulation). The mixing timescales are taken from RHD simulations \citep[][see section~\ref{s:RHD_clouds} below for details]{Freytag2010}. For the Allard et al. (2012a,b)\nocite{Allard2012a,Allard2012b}  pre-release of the BT-Settl models, the cloud model was improved by a dynamical determination of the supersaturation --- the ratio of the saturation vapor pressure to the compound local gas pressure ($P_g(t)/P_{sv}/$). In the current pre-release the cloud model is further improved by the implementation of a grain-size-dependent forward scattering, and by accounting for nucleation based on cosmic rays studies (Tanaka 2005). This latter change allows the cloud model to limit its refractory element depletion and form relatively more dust grains in higher atmospheric layers.

\begin{figure*}[!ht]
\includegraphics[width=6cm]{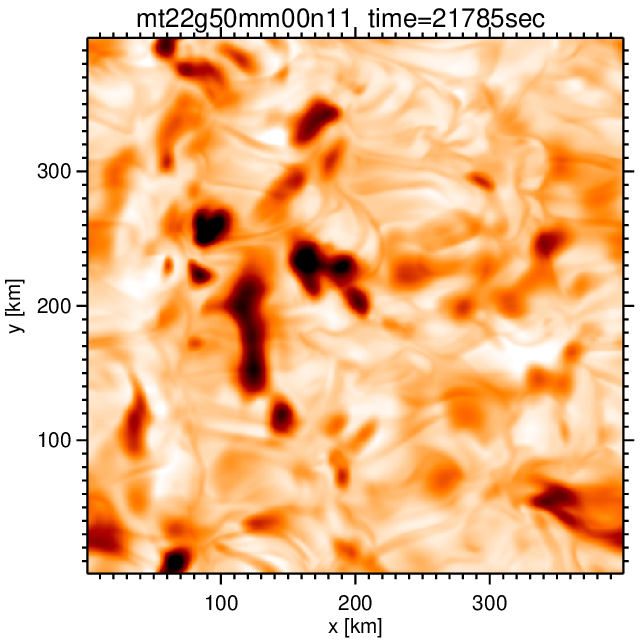}
\includegraphics[width=6cm]{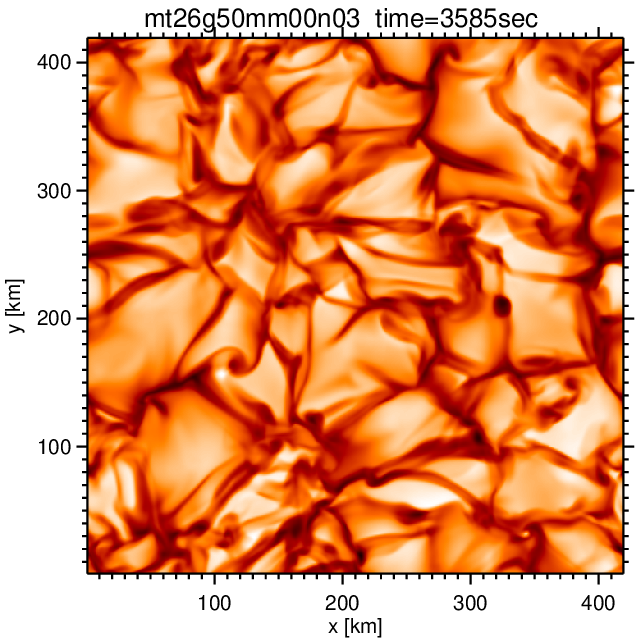}
\includegraphics[width=6cm]{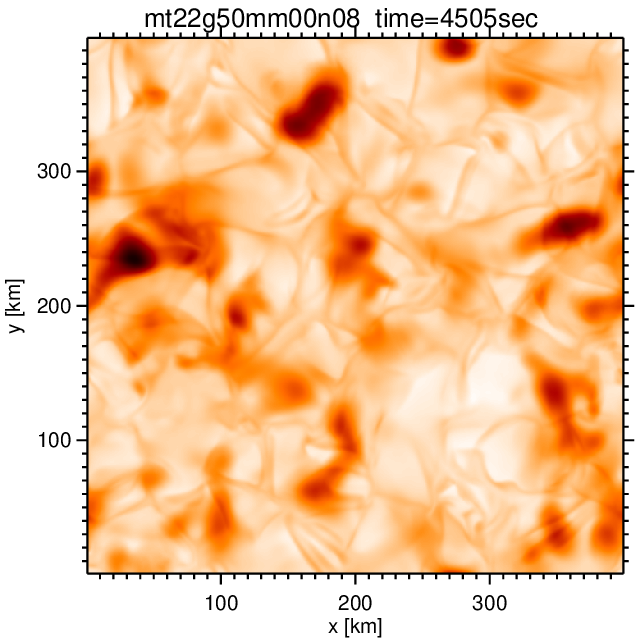}
\includegraphics[width=6cm]{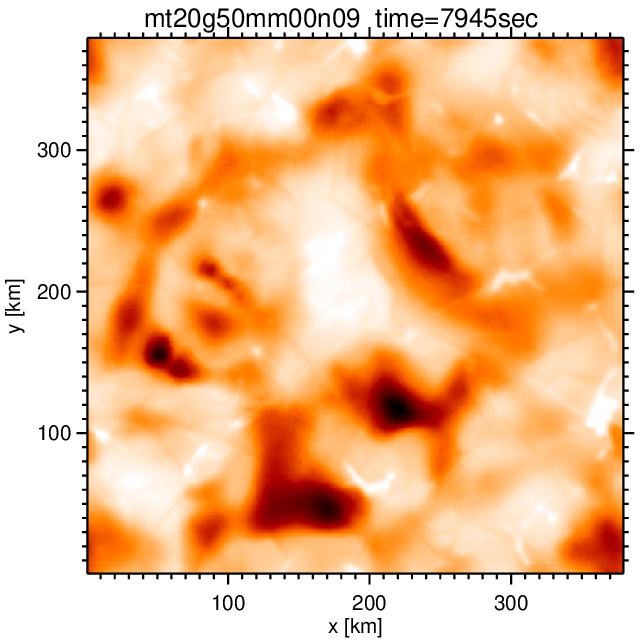}
\includegraphics[width=6cm]{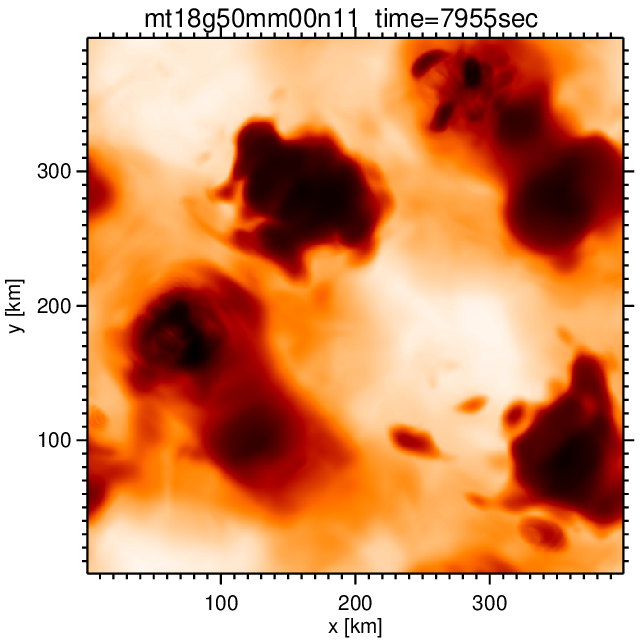}
\hspace{+1.4cm}
\includegraphics[width=6cm]{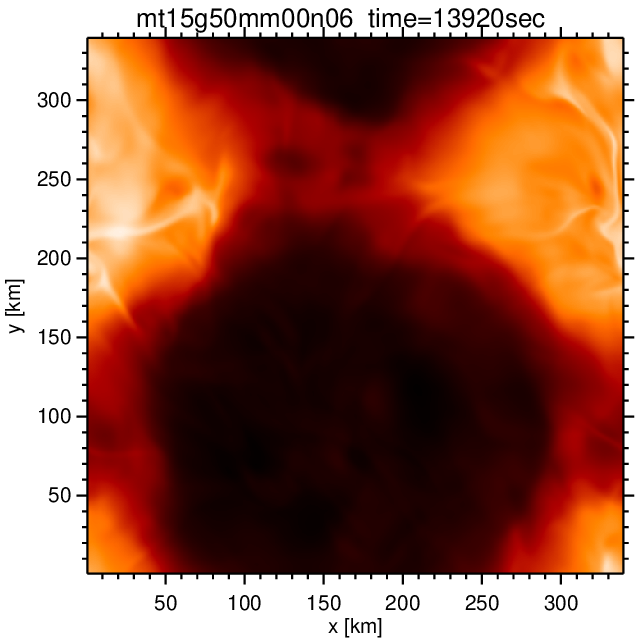}

\caption{\footnotesize 3D RHD simulations using {\sl\bf CO5BOLD}
    \cite[]{Freytag2010} of a small box of -- from top to
	right to bottom right -- 2600K, 2200K, 2000K, 1800K and 1500K
	atmospheres of $\mathrm{log}\,g$=5.0, solar metallicity.
        The models have no rotation, except for
	the 2200K case shown at the top left using an unrealistically small rotational period of 15
	minutes to highlight the -- otherwise negligible -- effects on the surface convection cells
        and the atmopheric wave pattern.}
   \label{f:3DHRD}
 \end{figure*}

The cloud model is solved from the innermost to the outermost atmospheric layer, depleting the gas composition due to sedimentation gradually from the bottom to the top of the atmosphere. This model obtains both the density distribution of grains and the grain average size per layer, and includes 55 types of grains. We do not assume a seed composition since our cloud model modifies the equilibrium chemistry iteratively to the nucleation rate limit at each atmospheric layers. The cloud composition in the photospheric layers varies with spectral type from zirconium oxide (ZrO$_2$) and refractory ceramics (perovskite and corundum; CaTiO$_3$, Al$_2$O$_3$) in late-M dwarfs, to silicates (enstatite, forsterite, etc.) in early-L dwarfs, and to salts (CsCl, RbCl, NaCl) and ices (H$_2$O, NH$_3$, NH$_4$SH) in late-T and Y dwarfs. This method is coherent with the observed weakening and vanishing of TiO and VO molecular bands (via CaTiO$_3$, TiO$_2$, and VO$_2$ grains) from the optical spectra of late M and L dwarfs, revealing CrH and FeH bands otherwise hidden by the molecular pseudo-continuum, and the resonance doublets of alkali transitions which are only condensing onto salt grains in late-T dwarfs.  We solve the Mie equation for spherical grains using complex refraction index of materials as a function of wavelength compiled by the Astrophysikalisches Institut und Universit\"ats-Sternwarte in Jena.

The results are shown in Fig.~\ref{f:allard_f4_Teff-J-Ks} where the new BT-Settl model atmospheres colors are interpolated onto the published theoretical isochrones \citep{Baraffe03}. New interior and evolution models consistent with the BT-Settl model atmospheres are in development.

\section{Cloud Formation in CO5BOLD}\label{s:RHD_clouds}

Cloud formation is included in the CO5BOLD simulations of VLMs and
brown dwarfs \citep{Freytag2010} by assuming a mass density of
forsterite dust ``monomers'' initially set to its maximum abundance at solar
metallicity, i.e. assuming no settling. Evaporation
is accounted for by applying a rate determined from the forsterite
saturation vapor curve. Grain growth by condensation and sedimentation 
are accounted for using the formulations described by \cite{Rossow1978}.  
The grains are assumed to have sizes up to $1~{\mu}$m and the corresponding 
geometric cross-section is added to the opacity bins on-the-fly.
Fig.~\ref{f:3DHRD} shows local 3D "box-in-a-star" type RHD
simulations using CO5BOLD and the same cloud model, where the
effects of cloud formation (restricted to forsterite formation) and
Coriolis effects (in the top left case) are investigated.  Here, the
centrifugal force has been neglected since over the box size the
centrifugal force would simply translate into a reduced gravity of the
simulation. The Coriolis force is exaggerated to explore
the maximum possible effect of rotation locally at the surface.

Global simulations of quasi-hydrostatic M, L and T dwarfs that can
resolve the convection and cloud formation processes, must be
significantly scaled down ($\approx 20$ so far) in radius compared to
the real object.  Conclusions from these RHD simulations must
therefore be wisely drawn from both types of local and global
simulations to learn about the cloud surface coverage of across the
M-L-T transition.  The preliminary conclusions are that the rotation
does not affect remarkably the convective shape and size of convective
\emph{surface} cells, the related gravity waves and the cloud formation process at
the surface.  We know also from several global circulation simulations
of rotating planets (periods around 3 days for tidally locked for
Jupiter around solar type stars) that strong winds and currents are
generated on larger scales at the surface.  These large scale features
are believed to be responsible for the distribution of clouds on the surface
of Jupiter.  The higher rotation velocities of brown dwarfs can
therefore similarly and easily cause comparable patterns at their
surface. However, it is not yet clear how patchy cloud coverage and
variability can occur. Corresponding global simulations are being
developed.

\section{Summary and Prospectives}

We showed that M dwarfs can be modeled adequately using up-to-date opacities and the revised solar abundances by \cite{Caffau2011} which
preserve the agreement with the results from solar astero-se\"\i smology \citep{Antia2011}. We also showed that VLMs and brown dwarfs allow to study the process of cloud formation in a relatively simple context where irradiation and ground interaction effects important for planets are not present. This allows to identify the fundamental mechanism of cloud formation and the determination of the velocity field by RHD simulations \citep{Freytag2010}. 

We have compared the behavior of the recently published model atmospheres from various authors across the M-L-T spectral 
transition from M dwarfs through L type and T type brown dwarfs and confronted them to constraints.  If the onset of dust formation
is occurring below \teff\ = 2900\,K, the greenhouse or line blanketing effects of dust cloud formation impact strongly ($J-K_s < 2.0$) 
the near-infrared SED of late-M and early L-type atmospheres for $1300 <$ \teff\ $< 2500$\,K.  The BT-Settl models by  
are the only models to span the entire regime from stars to planetary mass objects (300\,K$<$ \teff\ $< 70,000$\,K). In the M dwarf range,  the results appear to favor the BT-Settl based on the \cite{Caffau2011} solar abundances versus MARCS and ATLAS~9 models based on other values.   In the M dwarfs range, the BT-Settl models show an unprecedented fit quality, even in the $V$ bandpass despite still missing or incomplete opacities (in particular for the electronic bands of CaOH, and missing opacities in the corresponding MgH B-X system bands), above 2500\,K. In the brown dwarf (and planetary) regime, on the other hand, the unified cloud model by \cite{tsuji02} succeeds in reproducing the constraints, while the BT-Settl models also show 
a plausible transition.  No models succeed in reproducing perfectly the M-L transition between 2500 and 2000\,K at this stage. This \teff\ range 
is similar to that of young (directly observable by imaging) and strongly irradiated planets (Hot Jupiters). However, \cite{Bonnefoy2010,Bonnefoy2013}  obtain reasonable parameters and fit quality to the colors and SINFONI near-infrared integral field spectra of giant planets.  

Possible further development of the cloud model could lie in an improvement of the treatment of coagulation and of the grain opacities in general (e.g. porosity, non-sphericity, mixed grain composition rather than adding the contribution of pure grain species, etc.). 

In this paper, we have presented the first preliminary global RHD simulations of an M dwarf in presence of rotation.
This simulation is precise enough in the interior while able to resolve the atmospheric convective cells at the surface.
The strong molecular and atomic resonance line opacities prevailing in
M dwarf atmospheres lead to a control of their cooling evolution by
the atmosphere, such that interior and evolution codes must consider
the atmospheric structure as their boundary condition \citep{BCAH98}.
This requires the computation of large and fine
($\Delta$\teff $= 100$~K; $\Delta$log$g \le 0.5$~dex)
model atmosphere grids which will
however be difficult to be achieve by computationally expensive RHD codes like CO5BOLD. 
Implicit hydrodynamical simulations more adequate to address the interior convection conditions 
are being developed \citep{Viallet2011}. 
But these simulations will have to resolve the atmosphere and treat the radiative transfer 
and the surface cooling adequately. RHD simulations do not cover 
evolution timescales and cannot replace classical stellar evolution and atmosphere models. 
In the meanwhile, therefore, classical 1D static interior models are being developed using the new BT-Settl model atmospheres
as surface boundary conditions, and will be published shortly.

Global RHD simulation which account for rotation are needed to resolve the cloud surface distribution and induced potential variability. In principle, it is only a small step to go from the global RHD simulations of an M dwarf to global simulations of late-type M dwarfs, brown dwarfs, and gas giant planets that account for dust cloud formation and rotation. Such simulations are under development and to be expected for the end of 2013.

M dwarfs are magnetically active with possibly an important magnetic spot coverage due to their more concentrated magnetic field lines (same magnetic field strengths over smaller radii) than solar type stars. Magnetic spots are due to the local suppression of convection by the magnetic field lines emerging at these points of the stellar surface.  In the Sun, spots are showing an important suppression of the local convection, causing an important cooling (almost half the solar \teff). The suppression of convection efficiency by rotation and magnetic field is also known to lead to a slowed down contraction during their evolution, and to larger radii then predicted by classical interior models for M dwarfs with strong magnetic fields in close-binary systems \citep{Chabrier2007}. Global Magneto RHD (MRHD) simulations which account for rotation are needed to identify the mixing length to use in classical models to compensate for this suppression. The non-magnetic global RHD models presented in this paper already show that rotation alone can have a significant impact on the convective efficiency and the stellar radius. MRHD simulations are being developed also using CO5BOLD \citep[][Freytag et al. 2012,Wedemeyer-B\"ohm et al. 2012]{Nutto2012}\nocite{Freytag2012,Wedemeyer2012}. The CO5BOLD MRHD simulations have not yet been applied to global M dwarf models (to investigate the dynamo mechanism and the suppression of convection by magnetic fields in the interior, or to simulate magnetic spots). Fortunately however, we do not expect effects as important as for solar spots since M dwarf model atmospheres have proven to be
insensitive to the value of the mixing length \citep{ARAA97} unless of course if convection is completely suppressed at depth over the
stellar disk. The BT-Settl model atmospheres and synthetic spectra can therefore be used safely to determine the parameters of moderately active VLMs, 
brown dwarfs, and planetary mass objects. The BT-Settl synthetic colors and spectra are distributed via the \texttt{PHOENIX} web simulator\footnote{http://phoenix.ens-lyon.fr/simulator}.\\


\begin{acknowledgements}
The research leading to these results has received funding from the 
French ``Agence Nationale de la Recherche'' (ANR), the ``Programme 
National de Physique Stellaire'' (PNPS) of CNRS (INSU), and the
European Research Council under the European Community's Seventh 
Framework Programme (FP7/2007-2013 Grant Agreement no. 247060).
It was also conducted within the Lyon Institute of Origins under 
grant ANR-10-LABX-66. 


\end{acknowledgements}

\bibliography{allard_f,allard_f2,ref,aa_redsg}

\end{document}